# High-speed double layer graphene electro-absorption modulator on SOI waveguide

**Marco A. Giambra[1], Vito Sorianello[1], Vaidotas Miseikis[1,2], Simone Marconi[1,3], Alberto Montanaro[1], Paola Galli[4], Sergio Pezzini[2,5], Camilla Coletti[2,5,*], and Marco Romagnoli[1,**]**

[1]*Photonic Networks and Technologies Lab – CNIT, Via G. Moruzzi 1, 56124 Pisa (Italy)*
[2]*Graphene Labs, Istituto Italiano di Tecnologia, Via Morego 30, 16163 Genova (Italy)*
[3]*Tecip Institute – Scuola Superiore Sant'Anna, Via G. Moruzzi 1, 56124 Pisa (Italy)*
[4]*Nokia System Networks Italia, via Energy Park 14, 20871 Vimercate, (Italy)*
[5]*Center for Nanotechnology Innovation @NEST - Istituto Italiano di Tecnologia, Piazza San Silvestro 12, I-56127 Pisa (Italy)*
[*]*camilla.coletti@iit.it*
[**]*marco.romagnoli@cnit.it*

**Abstract:** We report on a C-band double layer graphene electro-absorption modulator on a passive SOI platform showing 29GHz 3dB-bandwith and NRZ eye-diagrams extinction ratios ranging from 1.7 dB at 10 Gb/s to 1.3 dB at 50 Gb/s. Such high modulation speed is achieved thanks to the quality of the CVD pre-patterned single crystal growth and transfer on wafer method that permitted the integration of high-quality scalable graphene and low contact resistance. By demonstrating this high-speed CVD graphene EAM modulator integrated on Si photonics and the scalable approach, we are confident that graphene can satisfy the main requirements to be a competitive technology for photonics.

## 1. Introduction

The exponential growth of data traffic, driven by the next 5G wireless communication [1] sets a continuous increase of demand for a technology matching the requirements in terms of bandwidth and power consumption. Integrated photonics is the most viable technology to target reliable and cost-efficient optical components. In particular, silicon (Si) photonics has been recognized as a convenient solution because of the well-established semiconductor technology platform allowing large volumes of production [2]. However, in the datacom and telecom roadmaps, the continuous bandwidth increase requires also a pathway of insertion losses, power consumption and cost reduction to enable next generation systems [3]. Within this scenario, in which conventional technologies may reach performance saturation, emerging materials with high electro-optical performance must be investigated.

Graphene is the ideal candidate because it offers electro-absorption [4] and electro-refraction effects [5] that can be used in modulators and switches [6], and photo-thermo-electric effect in detectors [7]. Graphene offers also the versatility to be integrated independently on different passive photonic platforms [3] such as Si, silicon nitride (SiN) and silica. However, the route to the demonstration of high Technology Readiness Level (TRL) devices and components requires the demonstration at laboratory level realized with wafer scale techniques, compatible with and easily transferrable to commercial foundries [3].

Graphene is a zero-bandgap material [8–10] allowing unique broadband optical absorption [11,12]. Its conductivity at optical frequencies is a complex quantity that affects both the optical absorption and the refractive index of the material [13,14], hence making graphene suitable for the fabrication of fundamental building blocks such as amplitude modulators, phase modulators and photodetectors [3,15]. To date, optical modulators based on a single layer of graphene integrated on top of Si doped waveguides have been successfully demonstrated up to 10 Gb/s for both EA and phase modulation [16,17]. Recently, single layer graphene EAMs on Si waveguide have been demonstrated up to 20 Gb/s NRZ [18]. These devices are based on a graphene-oxide-Si capacitor, in which the intentionally doped Si waveguide acts as a gate for the graphene sheet, ensuring low gate series resistance and high speed of operation [16]. Two single layers of graphene separated by 65nm of aluminum oxide ($Al_2O_3$) have been integrated on a SiN microring resonator to realize a wavelength-selective 22 Gb/s electro-optic modulator based on resonator loss modulation at critical coupling [19]. Dalir *et al* have demonstrated planar structure modulators with double-layer graphene with a 35 GHz optical bandwidth, however, high contact resistance and high driving voltage have prevented the measurement of an eye diagram for high speed data communications [20]. To date, graphene-based optical interconnect at 50Gb/s rate has been shown only on the detector side [21], while such high-speed data-rate broadband optical modulation has been hindered by graphene growth and transfer processing, high contact resistance or wavelength-dependent device design [19].

In this work we show for the first time a broadband high-speed graphene-based EAM with a 120μm long double layer geometry (DLG) and integrated on straight Si passive waveguide with 29 GHz electro-optical bandwidth, working up to 50 Gb/s non-return-to-zero (NRZ) modulation format. We rely on wafer scale seeded array fabrication method [22] and CMOS compatible processes to demonstrate how graphene technology can meet the highest performances required by the market.

## 2. Simulation

The EAM modulator is based on the modulation of the graphene absorption by electrical grating of the Fermi level ($E_F$) above the Pauli blocking condition (i.e. $E_F > \hbar\omega/2$ where $\hbar\omega$ is the photon energy at the specific wavelength) [11,13,23]. In the DLG geometry, both graphene layers act as gates, inducing a reciprocal tuning of $E_F$ [24]. The main advantages of this approach are the larger electro-absorption effect due to the presence of two graphene films, approximately a factor of two of enhancement, and the possibility to use simple undoped waveguides, enabling flexible integration onto any already existing platform [24]. Figure 1(a) shows a sketch and the cross section of the proposed device on the Si optical waveguide. The DLG capacitor is made of a graphene-SiN-graphene hetero-stack integrated on top of the Si waveguide to implement the EAM. Figure 1(b) shows a picture of the fabricated device at the optical microscope.

We used a 220 nm Si photonic platform, the waveguide is 450 nm wide, designed to support a single quasi-Transverse Electric (TE) in-plane polarized optical mode. The waveguide top cladding is thinned down to 10 nm to maximize the evanescent coupling of the optical mode with the DLG hetero-stack. The device cross-section has been optimized by means of optoelectronic simulations based on the graphene surface optical conductivity model [14]. We simulated the DLG EAM by using the complex surface conductivity of graphene in a commercial-grade mode solver assuming two possible values of the scattering time $\tau=1/\Gamma$, i.e 30 fs and 300 fs that corresponds to graphene with mobilities of ~660 and ~6600 $cm^2V^{-1}s^{-1}$ at $E_F = 0.4$ eV, respectively. The complex surface conductivity of graphene at optical frequencies $\sigma(\omega, \mu_c, \Gamma, T)$ ($\omega$ is the radian frequency, $\mu_c$ is the chemical potential or Fermi level $E_F$, $\Gamma$ is a phenomenological scattering rate assumed to be independent of energy, $T$ is the temperature) is described by the Kubo formula [14]:

$$\sigma(\omega,\mu_c,\Gamma,T) = \frac{jq^2(\omega-i2\Gamma)}{\pi\hbar^2}\left[\frac{1}{(\omega-i2\Gamma)^2}\int_0^\infty E\left(\frac{\partial f_d(E)}{\partial E} - \frac{\partial f_d(-E)}{\partial E}\right)dE - \int_0^\infty \frac{f_d(-E)-f_d(E)}{(\omega-i2\Gamma)^2 - 4(E/\hbar)^2}dE\right] \quad (1)$$

$$f_d(E) = \left(e^{(E-\mu_c)/k_BT}+1\right)^{-1} \quad (2)$$

where $q$ is the charge of the electron, $\hbar$ is the reduced Planck's constant, and $k_B$ is the Boltzmann's constant. The scattering rate $\Gamma$ stems from different scattering mechanisms, e.g. long-range charged impurity scattering, short-range surface-roughness or defect scattering, and affects the intra-band conductivity [14].

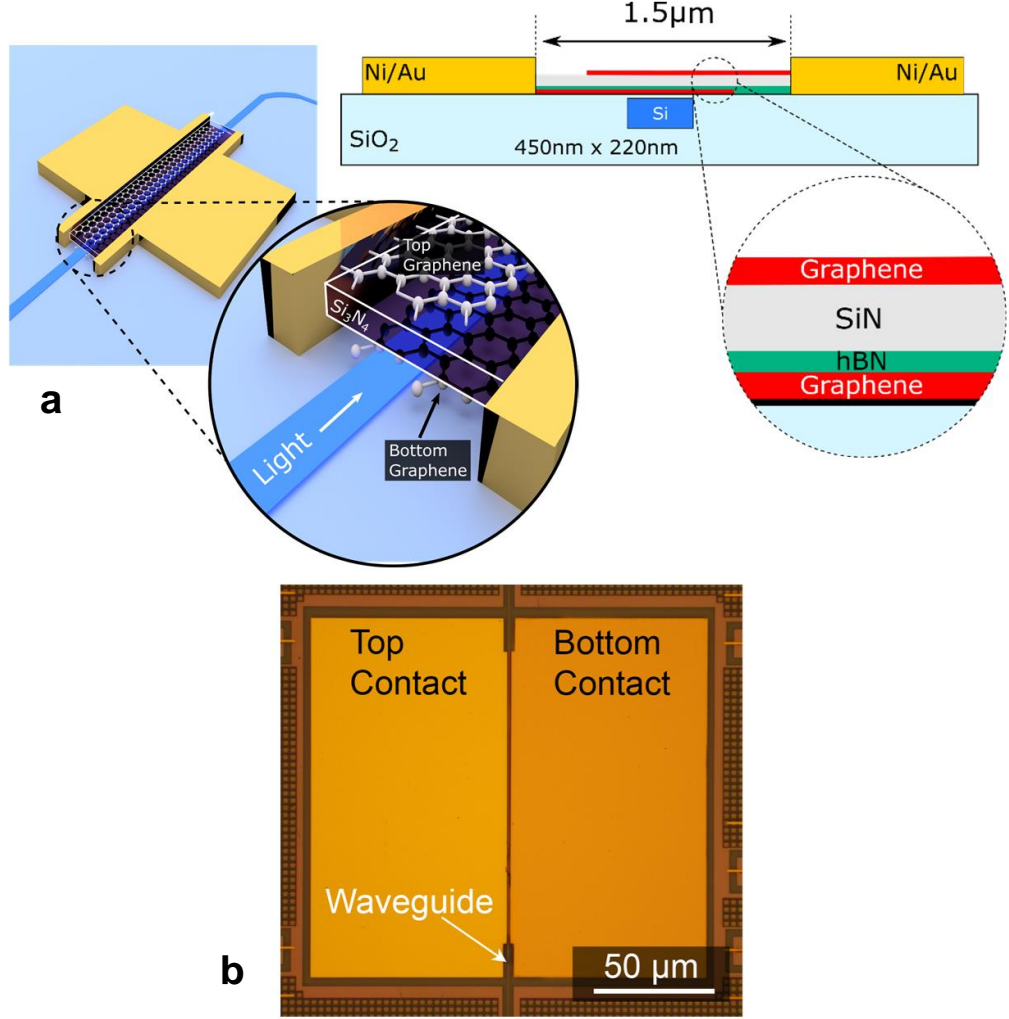

Fig. 1. (a) Schematic of the modulator consisting of a graphene-insulator-graphene capacitor integrated on a Si waveguide and cross-section of DLG EAM. The Si on insulator waveguide is a channel waveguide with 450nmx220nm core (blue) on 2µm buried oxide (BOX) and silica (SiO$_2$) lateral and top cladding (light blue). The top cladding is thinned and planarized to a final thickness of ~10nm on the top of the waveguide. The DLG hetero-stack is placed on top of the waveguide and consists of: a first single crystal graphene as bottom layer (red), a single layer of polycrystalline hexagonal Boron Nitride (hBN) (green), a ~20nm thin film of SiN (light grey), a second single crystal graphene as top layer (red). A stack of Ni(7nm)/Au(60nm) are used as top contacts on both graphene layers (yellow). The DLG stack is 650nm wide and 120µm long. The top and bottom graphene extends on opposite sides to the metal contacts that are separated by 1.5µm. (c) Optical micrograph of the fabricated device showing the waveguide and the Ni/Au contacts. Scale bar is 50µm.

We determined the complex effective index for the fundamental TE mode (Fig. 2(a)) of the waveguide and extracted the optical absorption at 1.55µm (Fig. 2(b)) as a function of the Fermi level and of the applied voltage calculated according to the following formula [25]:

$$|V - V_{Dirac}| = \frac{q\mu_c^2}{C_{ox}\pi(\hbar v_F)^2} + 2|\mu_c| \tag{3}$$

where $C_{ox}$ is the capacitance per unit length of the device, and $v_F = 9.5 \cdot 10^7$ cm/s is the Fermi velocity. The modulator efficiency and speed are highly dependent on the gating capacitor defined by the overlapping graphene layers, separated by the SiN dielectric film. A large capacitance is desirable to increase the modulation efficiency, i.e. low driving voltage and high Extinction Ratio (ER), while at the same time a small capacitance is needed for high speed operation. The optimally designed capacitor has a capacitance per unit length of ~1.7 fF/µm, with an overlap region of the two graphene layers of 650 nm and a SiN thickness of 20 nm with measured SiN dielectric constant of 6.

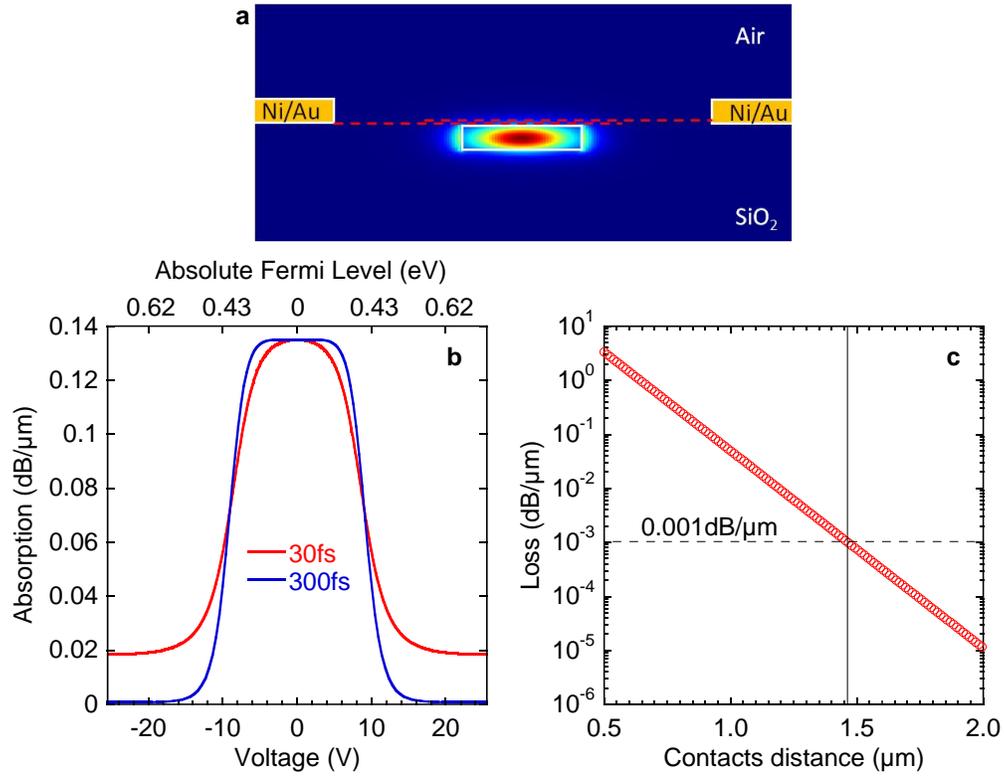

Fig. 2. (a). Transverse-Electric (TE) optical mode of the waveguide shown in fig. 1(b) calculated at 1.55µm wavelength with an optical mode solver. The dashed red lines represent the two graphene layers. (b). Optical propagation absorption per unit length of the optical mode shown in fig. 2(a) as a function of the voltage applied to the DLG capacitor (bottom x axis) and of the corresponding absolute value of the Graphene Fermi level (top x axis), for two different values of the scattering times associated to the mobility: 30fs (red line) (~660 cm$^2$V$^{-1}$s$^{-1}$ at $E_F = 0.4$eV), 300fs (blue line) (~6600 cm$^2$V$^{-1}$s$^{-1}$ at $E_F = 0.4$eV). (c) Optical propagation loss per unit length induced by metals as a function of the distance between metal contacts. We set a distance between the two metal contacts of 1.5µm to have a net loss contribution <0.1dB for a 120µm long device.

The maximum static ER per unit length depends on the mobility of the graphene, we calculated 0.137 dB/µm for a high mobility graphene (scattering time 300fs, i.e. ~6600 cm$^2$V$^{-1}$s$^{-1}$ at $E_F = 0.4$eV) when the voltage is changed from 0 to 20V, and 0.12 dB/µm for a low mobility graphene (scattering time 30fs, i.e. ~660 cm$^2$V$^{-1}$s$^{-1}$ at $E_F = 0.4$eV) in the same voltage range (see Fig. 2(b)). The distance between the metal contacts has been optimized to 1.5 µm in order to reduce the contribution of the ungated graphene regions, i.e. the regions of the two layers that don't overlap, to the series resistance, while keeping the distance between the metals and the waveguide sufficiently large to avoid extra optical losses (see Fig. 2(c)).

## 3. Fabrication

For the DLG EAM fabrication we used high quality single crystal graphene grown by chemical vapor deposition (CVD) on copper (Cu) foils by deterministic seeded growth [22], transferred on the Si waveguides by a semi-dry transfer technique demonstrated previously [22,26]. The used method allows a low level of metal contamination during the graphene delamination from the growth substrate as well as damage-free lamination onto the target substrate. A matrix of graphene crystals with 200 μm spacing and 150 μm crystal size was grown on Cu foils using chromium (Cr) nucleation seeds [22]. Cu foils were patterned with nucleation seeds using optical lithography and evaporation of 25 nm of Cr. Well-ordered arrays of graphene crystals were grown using a 4" BM Pro cold-wall reactor at a temperature of 1060 °C and a pressure of 25 mbar. Argon annealing and sample enclosure were used to maintain a low and well-controlled nucleation density [27]. The Cu foil with graphene crystals was spin-coated with a PMMA layer and an adhesive frame was attached to the perimeter of the sample. The graphene was then electrochemically delaminated from the Cu growth substrate by applying a voltage of -2.4 V with respect to a Cu counter-electrode. 1M NaOH solution was used as an electrolyte. After delamination the graphene array supported by the PMMA membrane and the frame was thoroughly rinsed in deionized water and aligned precisely to the target waveguides using a micrometric 4-axis stage. The lamination was performed at 90 °C. To improve adhesion, the sample was heated at 105 °C for 5 minutes, and the polymer support was then removed in acetone. The graphene was patterned using electron beam lithography (EBL) and reactive ion etching (RIE) (5 sccm oxygen and 80 sccm argon, 35 W power) using PMMA as an etch mask.

Low metal to graphene contact resistance is a key parameter to reach high speed of operation. In this work, we used top contacts made of a stack of Nickel (Ni) and Gold (Au). The top contacts were fabricated by EBL and thermal evaporation of 7 nm of Ni and 60 nm of Au, followed by lift-off in acetone. Following the fabrication of the bottom layer structure, the sample was coated with a protective layer of large-area poly-crystalline single-layer hexagonal Boron Nitride (hBN) provided by Graphene Laboratories Inc. hBN was supplied on Cu foil and semi-dry transfer procedure was used to laminate it on top of graphene. hBN provided significant protection to bottom layer graphene during the subsequent deposition of the SiN dielectric thin film. SiN was deposited using plasma-enhanced chemical vapor deposition (PECVD) at 350 °C. The top graphene layer was grown, transferred and patterned by using the same procedures utilized for the bottom layer. Crucially, a graphene matrix with an exactly-corresponding seeding pattern was used to ensure the precise overlap of graphene crystals for the bottom and top layer structures.

## 4. Material characterization

We used Raman spectroscopy [26] to characterize the two graphene crystals employed in the EAM. The Raman spectra were acquired using a Renishaw inVia system, with a 532 nm wavelength laser, laser power of 1 mW, and a 50x objective giving a laser spot diameter of ~2 μm. The spectrometer was calibrated using the Si peak at 521 cm$^{-1}$. For statistical analysis, we performed scanning Raman maps of the crystals, acquiring spectra every 10 μm and considering only regions where the two layers were not overlapping. The spectral peaks were fitted with single Lorentzian functions. Figure 3(a) shows two representative spectra for the bottom and top graphene layers, acquired after SiN deposition and after transfer onto SiN, respectively. Both spectra show the standard signatures of single-layer graphene, with sharp single-Lorentzian G and 2D peaks [28]. The negligible intensity of the D peak (expected at ~1350 cm$^{-1}$) indicates a very low density of defects. Nevertheless, the bottom graphene shows a blue-shifted 2D peak with reduced 2D to G peak intensity ratio, indicating increased doping

[29] and/or strain. The broader 2D peak of the bottom graphene is attributed to relevant strain fluctuations at the sub-μm scale [30] due to the plasma deposition of SiN. Considering that we are dealing with nearly defect-free single crystals, random strain fluctuations constitute a major limiting factor for the carriers' mobility [31], which, in turn, represents a key parameter for the performance of the EAM (see Fig. 2(a)). Therefore, in Fig. 3(b) we present a statistical analysis of the full width at half maximum (FWHM) of the 2D peaks, which assesses the electronic quality of the two graphene layers over device-size areas. The top layer shows an outstandingly low FWHM(2D) of $23 \pm 0.7$ cm$^{-1}$, a value comparable to what is commonly reported for pristine exfoliated graphene flakes on silica ($SiO_2$) [30]. The bottom layer, although subjected to the PECVD process, shows FWHM(2D) = $32.4 \pm 1.9$ cm$^{-1}$, lower than previously reported for CVD graphene adopted in high-performance EAM [17]. On average, we obtain position of the G peak Pos(G) = $1582.1 \pm 0.4$ cm$^{-1}$ ($1594.4 \pm 1.3$ cm$^{-1}$) with FWHM(G) = $13.9 \pm 1.1$ cm$^{-1}$ ($8.9 \pm 1.8$ cm$^{-1}$), and position of the 2D peak Pos(2D) = $2674 \pm 0.6$ cm$^{-1}$ ($2688.4 \pm 2.6$ cm$^{-1}$), for the top (bottom) layer. The average 2D to G area ratio is A(2D)/A(G) = $7.8 \pm 0.6$ ($6.6 \pm 1$), indicating p-type doping with $E_F$ ~ 125 meV (165 meV) and carrier density n ~$1.1 \times 10^{12}$ cm$^{-2}$ ($2 \times 10^{12}$ cm$^{-2}$), for the top (bottom) layer [29]. We also performed electrical characterization of the single crystal graphene on auxiliary test structure (transfer length methods (TLM) and Hall bars) with similar stack. We typically observed a maximum sheet resistance of ~1000 Ohm/sq for the bottom layer after the hBN and SiN deposition, and ~1500 Ohm/sq for the top layer.

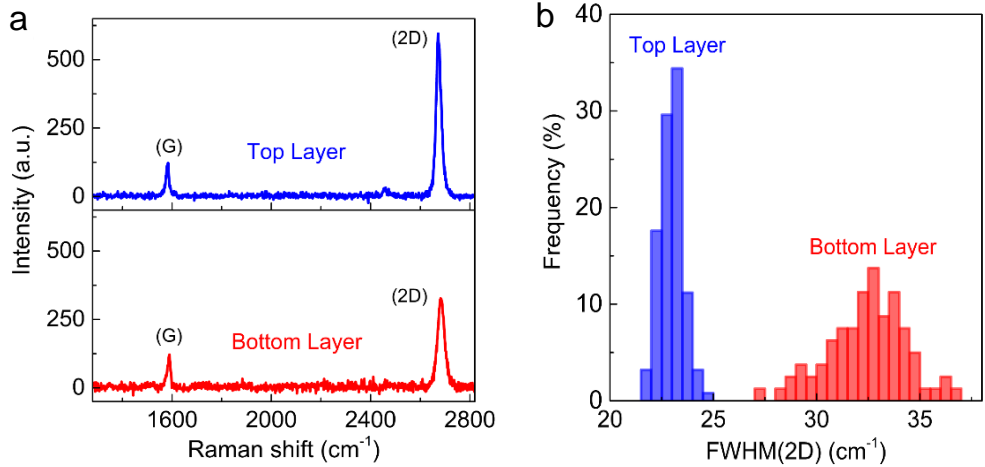

Fig. 3. (a) Raman spectra of the top (blue trace) and bottom (red trace) graphene layers. The spectra are normalized to the intensity of the G peak of the top layer. (b) Statistical distribution of FWHM(2D) for the top (blue histogram) and bottom (red histogram) layer.

## 5. Experimental results

The device was characterized in static and dynamic operations. We first characterized the DLG EAM transmission as a function of the bias applied to the capacitor. Two single-mode optical fibers (SMFs) with cleaved output surfaces were used to couple a tunable external cavity laser (ECL) fixed at 1550 nm into the chip by means of Si grating couplers. The input fiber was connected to the laser source with a polarization controller to match the polarization of the optical field at the input grating coupler. The output fiber was connected to a high-sensitivity power meter to measure the static characteristics. A signal-ground (SG) high-frequency probe was used to apply direct current (DC) bias to the DLG capacitor. Figure 4 shows the input/output transmission normalized to the insertion losses (IL) of the grating couplers (~ -11 dB measured on a reference waveguide) for the 120μm long DLG EAM.

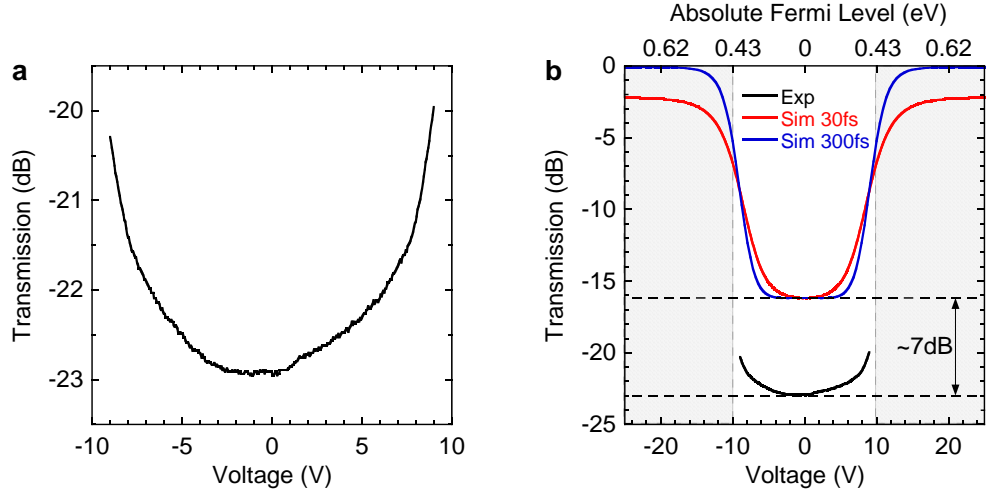

Fig. 4. (a) Optical transmission at 1550nm wavelength as a function of the voltage applied to the DLG capacitor. The transmission is normalized to the fiber to grating insertion loss (~ -11dB from the input to the output). (b) Comparison between the measured and simulated transmission as a function of the applied voltage and Fermi level. The dashed areas correspond to the bias regions where the fabricated dielectric suffers breakdown, i.e. above 10V corresponding to 5MV/cm for 20nm SiN.

The device exhibits a rather symmetric transmission suggesting a Dirac point close to zero voltage confirming a low intrinsic doping of the graphene films as expected by the Raman characterization. The EAM transmission can be varied from ~ -23 dB to ~ -20 dB by applying up to 9 V between the two layers of graphene, corresponding to a Fermi level of ~0.43eV on both layers, slightly above the Pauli blocking condition at 1550 nm. At higher biases the capacitor starts to exhibit a leakage current because of the limited breakdown field of the deposited SiN dielectric film. We have measured on a test sample a breakdown field of 5MV/cm, that is half the observed value reported in literature for an analogous process [32]. For this reason, we have limited the bias voltage to the safe region shown in Fig. 4(b), which shows a comparison between the simulations and experimental results. The obtained extinction ratio is mainly limited by the dielectric strength of the SiN spacer, which affects also the IL of the device. In fact, the maximum of the transmission is ~ -20dB at 9V (Fermi level of ~0.43eV), i.e. much smaller than expected (-2.2dB for the low mobility material, red curve in fig. 4(b)), because we did not bias the device in the low absorbing region (Fermi level >0.5eV).

Moreover, we measured ~7dB of extra loss in the maximum absorption region, which is attributed to increased propagation losses of the air cladded waveguide outside of the EAM due to impurities, mainly graphene and hBN residues, left on the waveguide during the fabrication of the DLG hetero-stack.

Improving the SiN dielectric strength and increasing the breakdown voltage, both the ER and IL can be significantly improved. For instance, with a state-of-the-art SiN film with breakdown voltage of 11MV/cm as reported in [32] we could easily drive the modulator in the transparency region up to 0.62eV. In this region, we would expect better agreement with simulations. i.e. IL < -2.2dB and the ER up to 16dB for a 120µm long DLG EAM (see Fig. 4(b)).

We characterized the electro-optical (EO) bandwidth of the DLG EAM using an electrical vector network analyzer (VNA). We used a bias-tee to combine the radio frequency (RF) signal from the VNA port 1 and the DC bias.

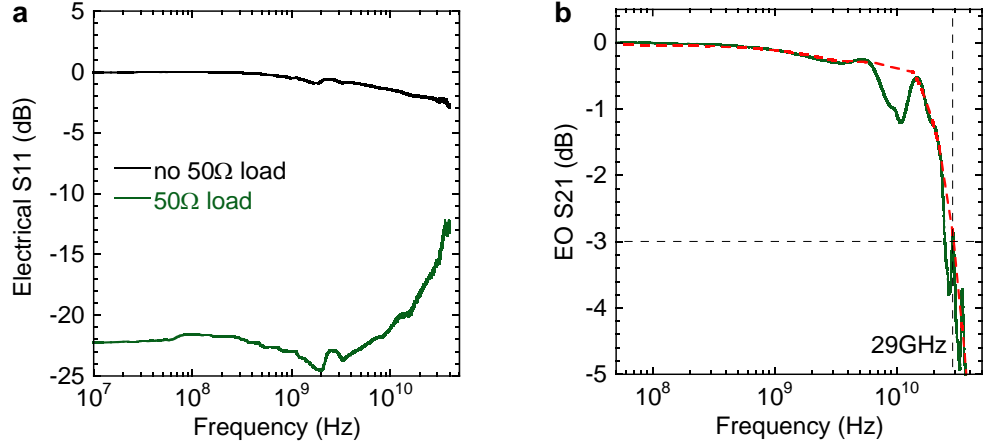

Fig. 5. (a) Electrical S11 parameter as a function of the frequency of the device with (green curve) and without (black curve) 50 Ω termination. (b) Electro-optical S21 bandwidth of the DLG EAM. The green line is the measured curve, the red dashed line the envelope fit. The measured -3dB bandwidth is ~29GHz mainly limited by the experimental setup loads.

The output of the bias-tee was connected through a 50 Ω matched high-frequency cable to the SG probe contacting the DLG EAM. A second SG probe was connected through a DC block to a high frequency 50 Ω termination load to reduce the reflections due to impedance mismatch between the VNA (50 Ω matched) and the lumped capacitive device. In fact, the 120 µm long EAM is sufficiently short to be a lumped device in the RF range under investigation (0-40GHz). In particular, the equivalent circuit is the series of the EAM capacitor (~204fF) and the resistance of to the ungated graphene regions and metal to graphene contacts (total resistance estimated <10 Ω). For this reason, the equivalent impedance of the device is much larger than the 50 Ω of the VNA, cables and SG probe. This is highlighted in Fig. 5(a) which shows the electrical S11 parameter of the device when measured with and without the 50 Ω load. When the device is probed without termination, we observe large reflections in the range of frequencies under investigation. When the device is shunted by the 50 Ohm load the reflections are reduced below -20dB. For these reasons, we measured the EO bandwidth of the device with the 50 Ω load termination shunted to the device to reduce the impairments due to the RF reflections. We set a DC bias of 8V and an RF power of 0 dBm. The light at the output of the DLG EAM was modulated by the RF signal from the VNA and collected by a low-noise, high-frequency photodetector connected to the VNA port 2. Fig. 5(b) shows the measured EO S21 as a function of the frequency, we measured 3dB roll-off frequency of ~29 GHz at 8V bias. The bandwidth is limited by the RC time constant due to the DLG capacitance and the equivalent resistance resulting from the contributions of the contact and sheet resistance of graphene, the 50 Ω output impedance of the VNA and the 50 Ω load shunted to the device. The measured bandwidth is close to the limit of ~31 GHz set by an RC time constant due to the only 50 Ω impedances, confirming a small contribution from the contact resistance of the device, measured to be around 300 Ω·µm at 0.2 eV doping on dedicated TLM and Hall bar test structures, and a nearly purely capacitive contribution of the two graphene layers. The measured EO bandwidth is compatible with very high-speed modulation.

We tested the EAM with a non-return-to-zero (NRZ) driving signal. The signals for the eye diagram were generated by a pattern generator (PG) and collected with a digital sampling oscilloscope (eye diagram). The PG provided the $2^{31}-1$ PRBS at different data rates up to 50 Gb/s. The electrical signal was sent to the DLG EAM through the RF cable and bias-tee. We used an erbium doped fiber amplifier (EDFA) to amplify the light at the input of the DLG EAM and compensate for the insertion loss of the device. The modulated light at the output of the EAM was collected by a low-noise, high-frequency photodetector connected to the oscilloscope to visualize the eye diagram of Fig. 6. The driving signal was 3.5 V peak to peak with 8V DC bias. The device was terminated with a 50 Ω load to avoid noise due to the impedance mismatch between the PG electrical output and the device. The bit rate of the signal was varied from 10 Gb/s up to 50 Gb/s (see Fig. 6).

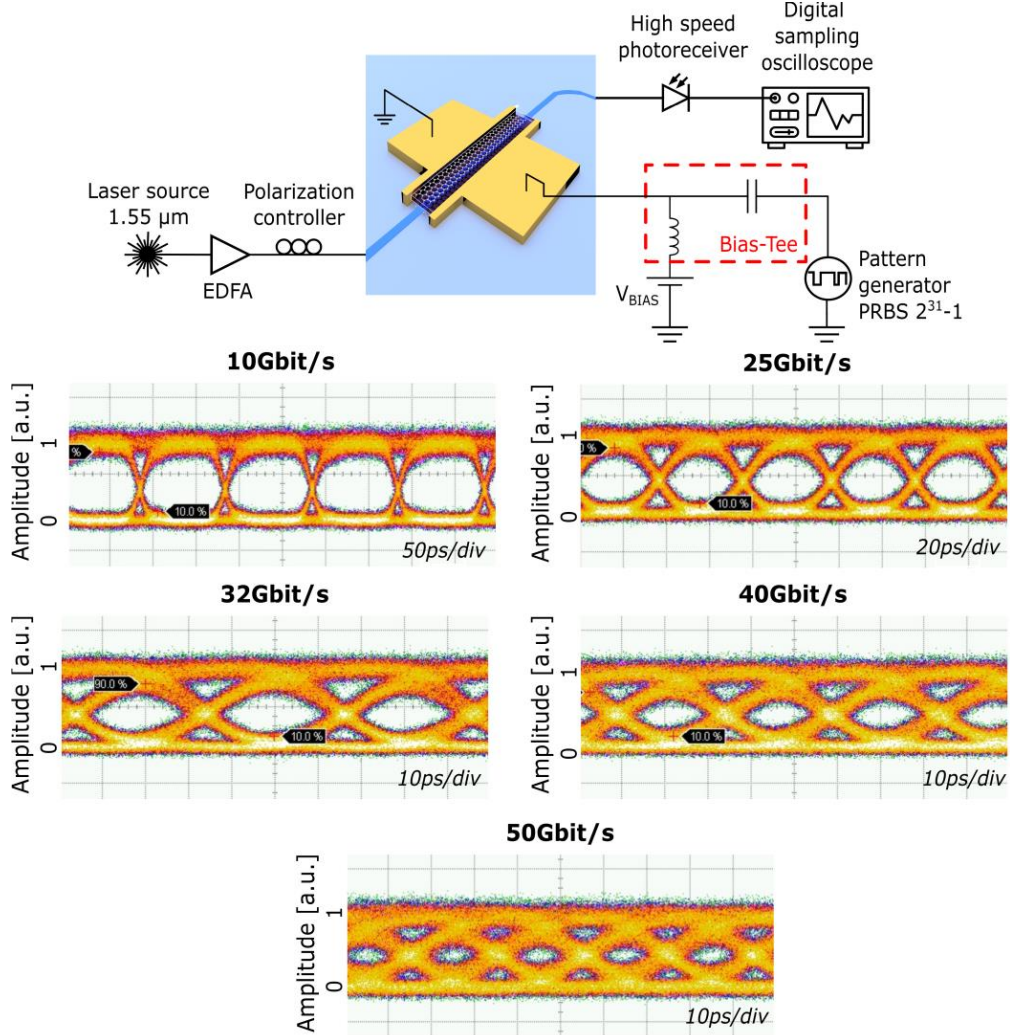

Fig. 6. Sketch of the setup and results of the NRZ modulation experiment. The light at 1.55μm is amplified by an EDFA to compensate for the device losses. A polarization controller is used to adjust the input polarization to match the polarization required by the input grating coupler. The NRZ signals were generated by a pattern generator and sent to the device through a bias-tee. The modulated light at the output of the EAM was collected by a low-noise, high-frequency photodetector connected to the oscilloscope to visualize the eye diagram. The reported eye diagrams are for $2^{31}$-1 PRBS at data rates of 10Gb/s, 25Gb/s, 32Gb/s, 40Gb/s and 50Gb/s.

The measured eye diagrams exhibit ER ranging from 1.7 dB at 10 Gb/s to 1.3 dB at 50 Gb/s. The poor ER is due to the limited bias that we could apply to the device, as discussed above, which set an operating point close to the Pauli blocking condition where the change in absorption is limited (see Fig. 4(b)). Moreover, the 50 Ω load termination, while improving the signal quality, reduces further the maximum extinction because the RF power driving the modulator is reduced by the shunt load.

## 6. Conclusions

In summary, an ultrafast DLG EAM has been demonstrated on a Si photonic waveguide. The device exhibits a maximum EO bandwidth of ~29 GHz. The EAM is capable of operating with NRZ $2^{31}$-1 PRBS signals up to 50 Gb/s data transmission rate. These results represent a major milestone in the development of next generation graphene based integrated photonics for datacom and telecom applications.

## Funding

European Union's Horizon 2020 research and innovation programme, GrapheneCore2 (grant agreement No 785219).

## Acknowledgments

We acknowledge IMEC for the provision of Si waveguides samples.